\documentstyle[preprint,aps,epsf]{revtex}

\begin{document}

\hyphenation{color-singlet color-exchange}

\title{Jet Production via Strongly-Interacting Color-Singlet Exchange
in $p\bar{p}$ Collisions}

% LIST_OF_AUTHORS.TEX                 09/07/95
%
\author{
S.~Abachi,$^{12}$
B.~Abbott,$^{34}$
M.~Abolins,$^{23}$
B.S.~Acharya,$^{42}$
I.~Adam,$^{10}$
D.L.~Adams,$^{35}$
M.~Adams,$^{15}$
S.~Ahn,$^{12}$
H.~Aihara,$^{20}$
J.~Alitti,$^{38}$
% ...
\\
\vskip 0.50cm
\centerline{(D\O\ Collaboration)}
\vskip 0.50cm
}
\address{
\centerline{$^{1}$Universidad de los Andes, Bogot\'{a}, Colombia}
\centerline{$^{2}$University of Arizona, Tucson, Arizona 85721}
\centerline{$^{3}$Brookhaven National Laboratory, Upton, New York 11973}
\centerline{$^{4}$Brown University, Providence, Rhode Island 02912}
\centerline{$^{5}$University of California, Davis, California 95616}
\centerline{$^{6}$University of California, Irvine, California 92717}
\centerline{$^{7}$University of California, Riverside, California 92521}
\centerline{$^{8}$LAFEX, Centro Brasileiro de Pesquisas F{\'\i}sicas,
                  Rio de Janeiro, Brazil}
\centerline{$^{9}$CINVESTAV, Mexico City, Mexico}
\centerline{$^{10}$Columbia University, New York, New York 10027}
\centerline{$^{11}$Delhi University, Delhi, India 110007}
\centerline{$^{12}$Fermi National Accelerator Laboratory, Batavia,
                   Illinois 60510}
\centerline{$^{13}$Florida State University, Tallahassee, Florida 32306}
\centerline{$^{14}$University of Hawaii, Honolulu, Hawaii 96822}
\centerline{$^{15}$University of Illinois at Chicago, Chicago, Illinois 60607}
\centerline{$^{16}$Indiana University, Bloomington, Indiana 47405}
\centerline{$^{17}$Iowa State University, Ames, Iowa 50011}
\centerline{$^{18}$Korea University, Seoul, Korea}
\centerline{$^{19}$Kyungsung University, Pusan, Korea}
\centerline{$^{20}$Lawrence Berkeley Laboratory and University of California,
                   Berkeley, California 94720}
\centerline{$^{21}$University of Maryland, College Park, Maryland 20742}
\centerline{$^{22}$University of Michigan, Ann Arbor, Michigan 48109}
\centerline{$^{23}$Michigan State University, East Lansing, Michigan 48824}
\centerline{$^{24}$Moscow State University, Moscow, Russia}
\centerline{$^{25}$University of Nebraska, Lincoln, Nebraska 68588}
\centerline{$^{26}$New York University, New York, New York 10003}
\centerline{$^{27}$Northeastern University, Boston, Massachusetts 02115}
\centerline{$^{28}$Northern Illinois University, DeKalb, Illinois 60115}
\centerline{$^{29}$Northwestern University, Evanston, Illinois 60208}
\centerline{$^{30}$University of Notre Dame, Notre Dame, Indiana 46556}
\centerline{$^{31}$University of Oklahoma, Norman, Oklahoma 73019}
\centerline{$^{32}$University of Panjab, Chandigarh 16-00-14, India}
\centerline{$^{33}$Institute for High Energy Physics, 142-284 Protvino, Russia}
\centerline{$^{34}$Purdue University, West Lafayette, Indiana 47907}
\centerline{$^{35}$Rice University, Houston, Texas 77251}
\centerline{$^{36}$Universidade Estadual do Rio de Janeiro, Brazil}
\centerline{$^{37}$University of Rochester, Rochester, New York 14627}
\centerline{$^{38}$CEA, DAPNIA/Service de Physique des Particules, CE-SACLAY,
                   France}
\centerline{$^{39}$Seoul National University, Seoul, Korea}
\centerline{$^{40}$State University of New York, Stony Brook, New York 11794}
\centerline{$^{41}$SSC Laboratory, Dallas, Texas 75237}
\centerline{$^{42}$Tata Institute of Fundamental Research,
                   Colaba, Bombay 400005, India}
\centerline{$^{43}$University of Texas, Arlington, Texas 76019}
\centerline{$^{44}$Texas A\&M University, College Station, Texas 77843}
}

\date{\today}
\maketitle

\begin{abstract}
A study of the particle multiplicity between jets with
large rapidity separation has been performed using
the  D\O\ detector at  the Fermilab Tevatron $p\bar{p}$ Collider operating
at $\sqrt{s}=1.8$\,TeV.
A significant excess of low-multiplicity events is observed
above the expectation for color-exchange processes.
The measured fractional excess is
$1.07 \pm 0.10({\rm stat})^{ + 0.25}_{- 0.13}({\rm syst})\%$,
which is consistent with a strongly-interacting color-singlet
(colorless) exchange process
and cannot be explained by electroweak exchange alone.
A lower limit of $0.80\%$ (95\% C.L.) is obtained on the fraction of
dijet events with color-singlet exchange, independent of
the rapidity gap survival probability.
\end{abstract}

\pacs{PACS numbers: 13.87.-a, 12.38.Qk, 13.85.Hd}

\draft

Jet production in hadron-hadron collisions
is typically associated with the exchange of
a quark or gluon between interacting partons.
In addition to these color-exchange processes,
the exchange of an electroweak color singlet (photon, $W$ or $Z$ boson) or
a strongly-interacting color singlet (such as two gluons in a colorless state),
can also produce jets.
Two jets separated by a rapidity gap, defined as a
region of rapidity
containing no final-state particles,
has been proposed as a signature for jet production via
the exchange of a color-singlet (colorless) object~\cite{Dok,BJ}.
Theoretical calculations have confirmed
that gluon radiation between scattered partons is
highly suppressed for color-singlet exchange relative to
color exchange~\cite{BF,DZ}.

The multiplicity of final-state particles
in the rapidity interval between
jets offers  a convenient way to distinguish
color-singlet exchange from color exchange.
Color-singlet exchange is expected to give a multiplicity near zero for
events that have no spectator interactions and
a minimum bias-like multiplicity distribution for events
that contain spectator interactions~\cite{BJ,BF}.
In contrast, color-exchange events are
expected to have a much higher mean multiplicity and to be
described by a negative binomial distribution (NBD) or
a sum of two NBD's (double NBD)~\cite{Pumplin,NB2}.
Low-multiplicity color-exchange events, which
are a background to color-singlet exchange, become
suppressed as the rapidity interval between the jets increases.
An excess of low-multiplicity events with respect to the
distribution for color exchange would %thus
indicate the presence of a color-singlet exchange process.

The magnitude of any observed excess
can be used to distinguish between a strongly-interacting or
purely electroweak color singlet.
The exchange of a two-gluon color-singlet, which has been proposed as
a model for the pomeron, is roughly estimated to account for
$10\%$ of the dijet cross section~\cite{BJ,UW,Duca}, while the
contribution from electroweak exchange is calculated to be about
$0.1\%$~\cite{UW}.
The survival probability ($S$) for
rapidity gaps to contain no particles from spectator interactions
is estimated to be 10--30\%~\cite{BJ,BF,GLM}.  Thus,
the fraction of dijet events with an observable rapidity gap
is expected to be about $1$--$3\%$
for two-gluon color-singlet exchange
and $0.01$--$0.03\%$ for electroweak exchange.

Although evidence exists for
color-singlet (pomeron) exchange
in single-diffractive jet events~\cite{UA8,HERA},
these events are typically produced with low momentum transfer,
$0\!<\!|t|\!<\!2$ $({\rm GeV}/c)^2$.
The D\O\ and CDF Collaborations have previously reported
the observation of  rapidity gaps in dijet events~\cite{PRL,CDF}, which,
in contrast to single-diffractive jet events,
have a rapidity gap between the jets and $|t|\!>\!900$\,$({\rm GeV}/c)^2$.
This letter presents a new analysis,
based on the same data as in Ref.~\cite{PRL},
that provides clear experimental evidence for jet production via
color-singlet exchange at a level inconsistent with an electroweak exchange
process alone.
In this work, pseudorapidity,
$\eta \equiv -\ln \tan (\theta /2)$, is used as an %good
approximation for true rapidity.

The D\O\ detector and trigger system are described elsewhere~\cite{NIM}.
The primary trigger used in the present analysis was previously discussed in
Ref.~\cite{PRL} and required ``opposite-side'' jets with a large
pseudorapidity separation.
In the offline analysis, the two leading $E_T$ (highest transverse energy)
jets are required to have
$E_T\!>\!30$\,GeV, $|\eta| \! > \!2$, and $\eta_1 \cdot \eta_2 <0$.
A cone algorithm with radius
${\cal R}\!=\!(\Delta\eta^2\!+\!\Delta\phi^2)^{\frac{1}{2}}\!=\!0.7$
is used for jet finding.
Events with more than one proton-antiproton interaction
are removed since extra interactions would obscure a color-singlet
signature and alter the multiplicity distribution.
This single interaction requirement yields a final data sample of 22,400
opposite-side jet events.

An independent trigger required events on the ``same-side'' of the detector
in pseudorapidity.  A final sample of
23,200 same-side jet events
is obtained by requiring  a single interaction and two jets with
$E_T\!>\!30$\,GeV, $|\eta|\!>\!2$, and $\eta_1 \cdot \eta_2\!>\!0$.
The same-side sample provides a qualitative measure of the
color-exchange background multiplicity in the central rapidity region
due to the color flow between the scattered and spectator partons.
Hard single diffraction, which
could produce a central rapidity gap
with two forward jets,
is highly suppressed by the trigger which required a
coincidence of hits between the forward and backward
luminosity counters ($1.9\!
\mathrel{\rlap{\raise 0.4ex \hbox{$<$}}{\lower 0.72ex \hbox{$\sim$}}} \!
|\eta| \! \mathrel{\rlap{\raise 0.4ex \hbox{$<$}}
{\lower 0.72ex \hbox{$\sim$}}} \! 4.3$).

The electromagnetic (EM) section of the calorimeter
is used as the primary means of measuring
the particle multiplicity.  % between jets.
The EM calorimeter has a low level of noise
and the ability to detect (with an energy-dependent efficiency)
both neutral and charged particles
for $|\eta|\!\mathrel{\rlap{\raise 0.4ex \hbox{$<$}}
{\lower 0.72ex \hbox{$\sim$}}}\!4$.
A particle is tagged by the deposition of more than
200 MeV transverse energy
in an EM calorimeter tower
($\Delta\eta\! \times\! \Delta\phi = 0.1 \!\times\! 0.1$).
The central drift chamber (CDC) is
efficient for detecting charged particles for
$|\eta| \! \mathrel{\rlap{\raise 0.4ex \hbox{$<$}}
{\lower 0.72ex \hbox{$\sim$}}} \! 1.3$ and
provides an independent measurement of the multiplicity.

\begin{figure}
\vspace{.5in}
\epsfxsize=6in
\leavevmode\epsffile[20 286 527 472]{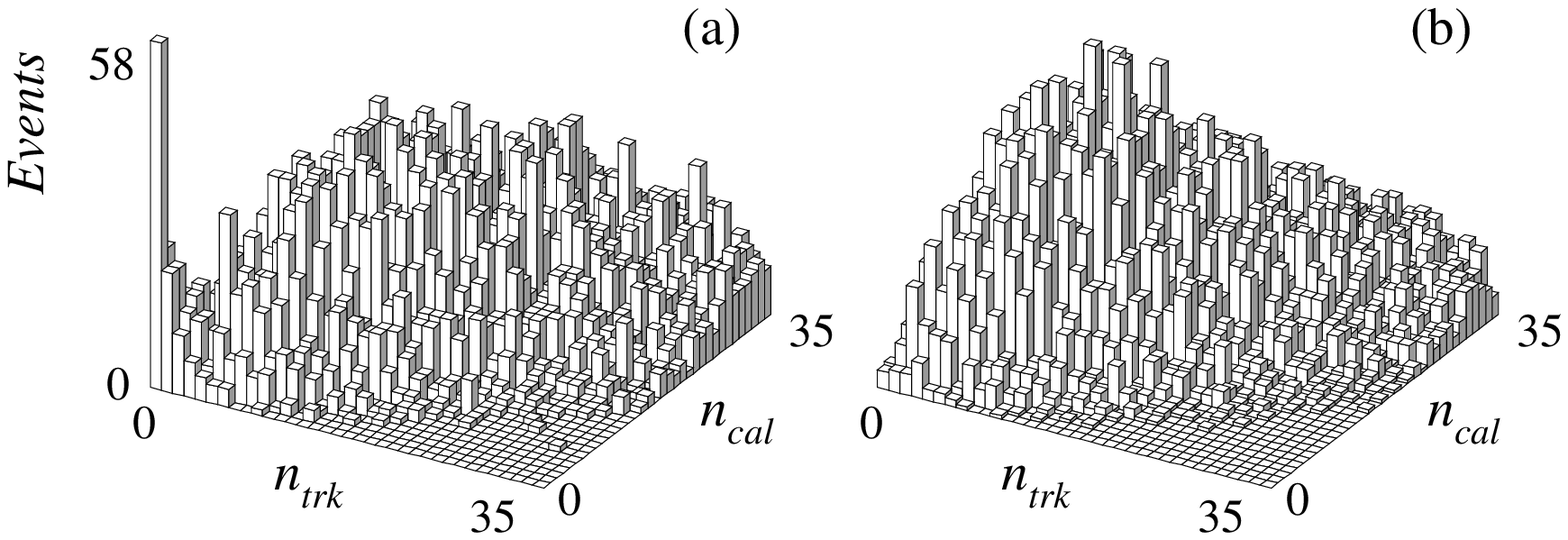}
\caption{The calorimeter tower multiplicity ($n_{\rm cal}$)
versus the charged track multiplicity ($n_{\rm trk}$) in the
pseudorapidity region $|\eta|\!<\!1.3$ for the (a) opposite-side and
(b) same-side samples as described in the text.}
\label{f:2D}
\end{figure}

Figure~\ref{f:2D} shows the number of EM calorimeter towers above threshold
($n_{\rm cal}$) versus the number of CDC tracks ($n_{\rm trk}$)
in the region $|\eta|\!<\!1.3$ for the
(a) opposite-side and (b) same-side samples.
The two distributions are similar in shape
except at very low multiplicities, where the opposite-side sample
has a striking excess of events,
consistent with a color-singlet exchange process.
For both samples, $n_{\rm cal}$ and $n_{\rm trk}$ are strongly correlated,
confirming that either can be used as a measure of particle multiplicity.

A model for the multiplicity in color-exchange events is necessary
to measure the low-multiplicity excess observed in the opposite-side data.
We use the double NBD,  which has
four parameters and a relative normalization,
to parametrize the color-exchange component
of the opposite-side multiplicity distribution between jets.

\begin{figure}
\vspace{.5in}
\epsfxsize=6in
\leavevmode\epsffile[15 266 527 527]{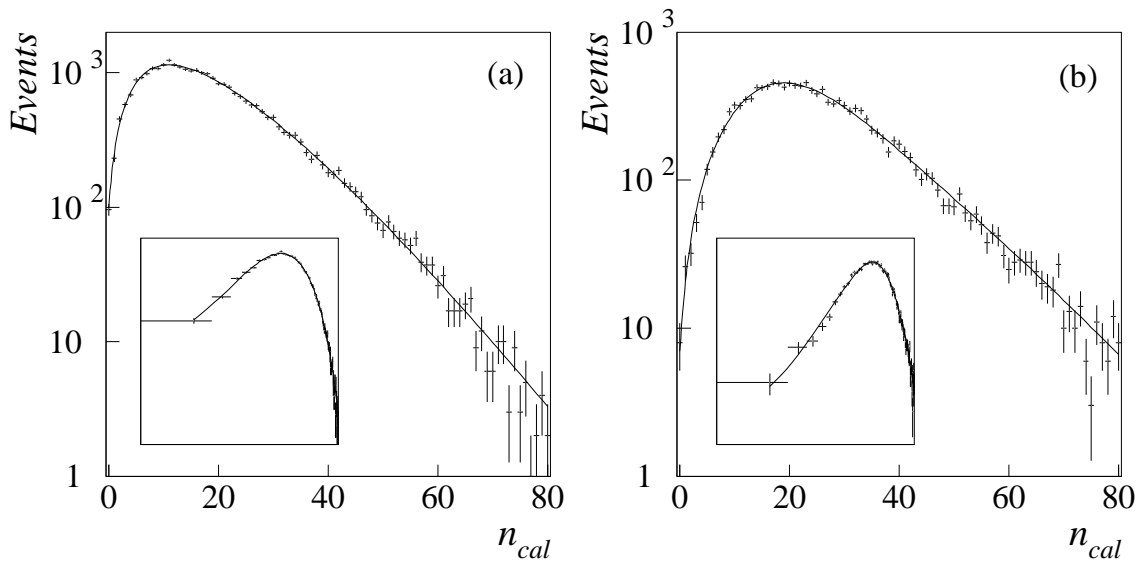}
\caption{The calorimeter tower multiplicity ($n_{\rm cal}$) for
the (a) same-side and
(b) color-singlet exchange suppressed data samples.
The solid lines are double NBD fits to the data.
The insets show each plot on a log-log scale.}
\label{f:back}
\end{figure}

This parametrization is supported by
color-exchange data and Monte Carlo  samples.
Figure~\ref{f:back}(a) shows
that the same-side $n_{\rm cal}$ distribution  for $|\eta|\!<\!1$
is well-parametrized  by  the double NBD
over the full range of multiplicity giving a
$\chi^2/df\,(\mbox{\rm degree of freedom})\!=\!0.9$.
Additional support is given by a
color-singlet-exchange-suppressed subset of the opposite-side data
obtained by demanding the presence of a third jet (with $E_T\!>\! 8$\,GeV)
between the two leading jets.
Figure~\ref{f:back}(b) shows the multiplicity
for these events in the pseudorapidity
region between the jet cone edges, excluding the
multiplicity of the third jet. %to avoid biases from this jet.
The double NBD fits the distribution reasonably well ($\chi^2/df\!=\!1.2$)
over the full range of multiplicity.
Monte Carlo color-exchange events, generated with
{\footnotesize HERWIG}~\cite{HERWIG} and {\footnotesize PYTHIA}~\cite{PYTHIA}
and passed through a simulation of the D\O\ detector~\cite{Pumplin,bob},
provide further support for the double NBD (not shown).
None of these color-exchange samples
have a significant excess of events at low multiplicity
from  physics processes or detector effects.

The low-multiplicity excess observed in Fig.~\ref{f:2D}(a)
is determined as follows.
First, a double NBD is fit to the opposite-side
calorimeter multiplicity distribution between the two leading jets
using a binned maximum likelihood method. %to determine the fit parameters.
The overall normalization is set to the number of events in the fitted region
and the relative normalization of the two NBD's is determined by minimizing
a modified Kolmogorov-Smirnov (K-S) statistic, which is
defined as the sum of the squared differences between the
cumulative distributions of the data and the fit.
Next, the starting bin ($n_0$) for the fit is incremented successively by one
and the remaining bins refit
until the $\chi^2/df$ in the first five bins of the fit
is less than 2.0 (i.e. C.L.\,$ >\! 7\%$).
The final fit is then extrapolated to zero multiplicity to determine
the excess above the expected color-exchange background for $n_{\rm cal}<n_0$.
Note that any
experimental effects that
produce a smearing of  zero multiplicity color-singlet events
to slightly higher multiplicities
are reduced by the integration over %a suitable range of
low-multiplicity bins.

\begin{figure}
\vspace{.5in}
\epsfxsize=6in
\leavevmode\epsffile[15 266 527 527]{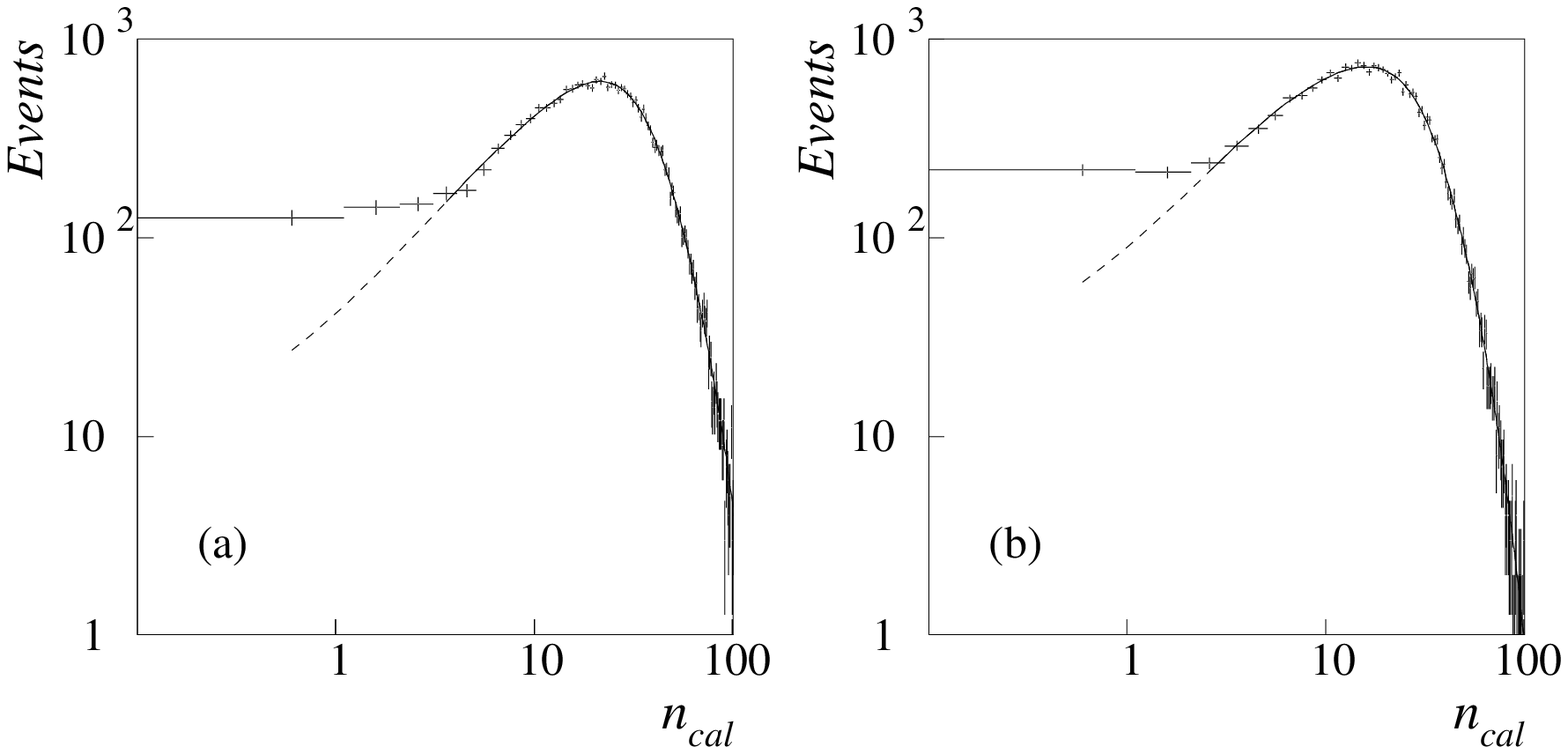}
\caption{The number of events versus $n_{\rm cal}$ (shifted up by
half a unit in multiplicity) is
plotted on a log-log scale
for all opposite-side jet events, where
(a) shows the multiplicity  between
the cone edges of the two leading jets and
(b) shows the multiplicity in the region $|\eta|\!<\!1$.
The solid lines represent the double NBD fits
for (a) $n_0\!=\!3$ and (b) $n_0\!=\!2$,
while the dashed lines show the extrapolation
of each fit to $n_{\rm cal}\!=\!0$.}
\label{f:OS}
\end{figure}

Figure~\ref{f:OS}(a) shows the $n_{\rm cal}$ distribution
between the cone edges of the two leading jets
for  the opposite-side data sample.
Also shown is the double NBD fit ($\chi^2/df\!=\!0.9$, $n_0\!=\!3$)
and its extrapolation to zero multiplicity.
In contrast with the color-exchange data and Monte Carlo samples,
the opposite-side sample has a significant excess
at low multiplicity.
The multiplicity in the central calorimeter
($|\eta| \! < \! 1$), a region well away from the
jet edges, is plotted in Fig.~\ref{f:OS}(b) for the same events.
Although the distribution has a lower mean multiplicity,
its shape (excluding low-multiplicity bins) is also well-fit by a
double NBD ($\chi^2/df\!=\!1.0$, $n_0\!=\!2$),
and a clear excess is again present at low multiplicity.
The excess, defined as the integrated difference between the data and the
double NBD fit in the extrapolated region,
is $225\pm 20$ and $237\pm 21$ events for
Figs.~\ref{f:OS}(a) and (b), respectively.

Using a single NBD instead of a double NBD to
fit the data in Figs.~\ref{f:OS}(a) and (b), as done in Ref.~\cite{GLASGOW},
gives a significantly larger excess of 339 and 421 events, respectively.
Although the single NBD fit gives a $\chi^2/df \! = \! 1.1$,
K-S tests show that the shape of the double NBD is
significantly favored over the single NBD.

The excess observed at low multiplicity is consistent with
the presence of color-singlet exchange, but
the fractional excess is a more relevant quantity for
theoretical comparisons.
The fractional excess $f_S$ is
defined as  the average excess above the double NBD fit
for starting points of $n_0$ and $n_0\!+\!1$
divided by the total number of events.
We obtain a fractional excess of $f_S \!=\! 1.07\%$
from the multiplicity distribution between the jet edges shown in
Fig.~\ref{f:OS}(a),
and a value of 1.14\%
using the distribution in the region $|\eta|\!<\!1$ from Fig.~\ref{f:OS}(b).

The exact fit parameters depend on the definition of particle multiplicity,
but the fractional excess is relatively independent of this definition.
Varying the calorimeter tower $E_T$ threshold
and refitting the multiplicity yields a consistent value of $f_S$.
Redefining a particle as a CDC track or a ``cluster'' of
neighboring calorimeter towers results in an excess that is 3--15\% greater.
We thus
assign a conservative $+15\%$ systematic error on $f_S$
due to the uncertainty in defining particle multiplicity.

Systematic effects from uncertainties in jet reconstruction,
energy scale, and acceptance
result in a $\pm 5\%$ uncertainty in the measured excess.
Uncertainty in the identification of single-interaction events
gives an additional error of $^{+14}_{-4}\%$.
%Effects from misidentification of
%events that have single proton-antiproton interactions
%give an additional uncertainty of $\pm 6\%$.
A systematic error of $\pm 10\%$ due to the uncertainty
in the background is determined by
varying the fit parameters to produce a one unit change in $\chi^2$
%(equivalent to $1\sigma$ variation of correlated parameters)
and remeasuring the excess.
Combining the above errors in quadrature
gives
$f_S \!=\!1.07 \pm 0.10({\rm stat})^{ + 0.25}_{- 0.13}({\rm syst})\%$.

Although a low-multiplicity excess is clearly present in
the opposite-side sample,
theoretical interpretation of $f_S$ is complicated by uncertainties in the
survival probability.
In contrast with our previous measurement which
used the number of events with zero multiplicity to place an upper limit on
the rapidity gap fraction
$((\sigma_{\rm singlet}/\sigma)
\times S<1.1\%\ \mbox{\rm at 95\% C.L.})$~\cite{PRL},
the measurement of
$f_S$ may include some portion of the color-singlet events
with low-multiplicity spectator interactions.
Nevertheless, the value of $f_S$ is consistent with the 1--3\%
expected for strongly-interacting color-singlet exchange.
%The value of the excess is used to
The measured value of  $f_S$ is used to
obtain a lower limit of $0.80\%$ (95\% C.L.) on
$\sigma_{\rm singlet}/\sigma$, the fraction of dijet events produced via
color-singlet exchange, independent of the actual value of the
survival probability.

The opposite-side data can also be used to exclude pure electroweak
exchange combined with color-exchange background as the source of the excess.
A {\footnotesize PYTHIA} Monte Carlo study
using simulated D\O\ jet acceptance and efficiency
gives a value of $0.09\%$ for the fraction of dijet events
with electroweak exchange, which
is comparable to the result in Ref.~\cite{UW} but includes
higher-order radiative corrections.
We assume a survival probability of $S\!=\!100\%$
to determine the maximum expected number of events from electroweak exchange,
and add this to the number of color-exchange background events
from the fit.
The  statistical and systematic uncertainty in the expected number of events
is then used to determine the probability for a fluctuation to
the number of observed events.
Using only the zero multiplicity bin,
the probability that the observed
excess in Fig.~\ref{f:OS}(a) is due to the combination of
electroweak and color exchange is less than $10^{-10}$.
Adding other low-multiplicity bins
further decreases this probability.

In conclusion, we have presented %experimental
evidence for strongly-interacting
color-singlet exchange from a study of opposite-side dijet events with
jet $E_T\!>\!30$\,GeV and $|\eta|\!>\!2$.
A striking enhancement of low-multiplicity events is observed
independent of the details of the color-exchange background
parametrization.
The double NBD parametrization provides a measurement
of the color-exchange background and confirms that
this background is small for the rapidity gap fraction previously
measured in Ref.~\cite{PRL}.
The fractional excess above the color-exchange background
is  found to be $1.07 \pm 0.10({\rm stat})^{ + 0.25}_{- 0.13}({\rm syst})\%$
which is consistent with the presence of  strongly-interacting color-singlet
exchange and inconsistent with electroweak exchange alone.

We thank the Fermilab Accelerator, Computing, and Research Divisions, and
the support staffs at the collaborating institutions for their contributions
to the success of this work.   We also acknowledge the support of the
U.S. Department of Energy,
the U.S. National Science Foundation,
the Commissariat \`a L'Energie Atomique in France,
the Ministry for Atomic Energy and the Ministry of Science and
Technology Policy in Russia,
CNPq in Brazil,
the Departments of Atomic Energy and Science and Education in India,
Colciencias in Colombia, CONACyT in Mexico,
the Ministry of Education, Research Foundation and KOSEF in Korea
and the A.P. Sloan Foundation.

%%%%%%%%FIG1

%%%%%%%%FIG2

%%%%%%%%FIG3


\begin{references}

% LIST_OF_VISITOR_ADDRESSES.TEX                            09/07/95
%
\bibitem[*]{conicet}
Visitor from CONICET, Argentina.

\bibitem[\dag]{buenosaires}
Visitor from Universidad de Buenos Aires, Argentina.

\bibitem[\ddag]{beijing}
Visitor from IHEP, Beijing, China.

\bibitem[\S]{ecuador}
Visitor from Univ. San Francisco de Quito, Ecuador.

\vskip 0.25cm


\bibitem{Dok} Yu.L. Dokshitzer, V.A. Khoze and S.I.
              Troian, {\it Proceedings of the 6th
              International Conference on Physics in Collisions} (1986), ed.
              M. Derrick (World Scientific, 1987).
\bibitem{BJ} J.D. Bjorken, Phys. Rev. D {\bf 47}, 101 (1992).
\bibitem{BF}  R.S. Fletcher and T. Stelzer, Phys. Rev. D {\bf 48}, 5162 (1993).
\bibitem{DZ} H. Chehime and D. Zeppenfeld, preprint MAD-PH-814 (1994).
\bibitem{Pumplin} J. Pumplin   Phys. Rev.  D {\bf 50}, 6811 (1994).
\bibitem{NB2}   I.M. Dremin, preprint FIAN TD-6, (1994).
\bibitem{UW} H.N. Chehime {\it et al.}, Phys. Lett. B {\bf 286}, 397 (1992).
\bibitem{Duca} V. Del Duca and W.K. Tang, {\it Proceedings of the 5th
              Blois Workshop on Elastic and Diffractive Scattering} (1993), ed.
               H.M.~Fried {\it et al.}, (World Scientific, 1994).
\bibitem{GLM} E. Gotsman, E.M. Levin and  U. Maor, Phys. Lett. B {\bf 309},
             199 (1993).
\bibitem{UA8} A. Brandt {\it et al}. (UA8 Collaboration), Phys. Lett. B
             {\bf 297}, 417 (1992).
\bibitem{HERA} M. Derrick {\it et al}. (Zeus Collaboration), Phys. Lett. B
             {\bf 332}, 228 (1994);
             Phys. Lett. B {\bf 346}, 399 (1995);\\
             T. Ahmed {\it et al.} (H1 Collaboration), Nucl. Phys.
             {\bf B435}, 3 (1995).
\bibitem{PRL} S. Abachi {\it et al.} (D\O\ Collaboration),
              Phys. Rev. Lett. {\bf 72}, 2332 (1994).
\bibitem{CDF}  F. Abe {\it et al.} (CDF Collaboration),
               Phys. Rev. Lett. {\bf 74}, 855 (1995).

\bibitem{NIM} S. Abachi {\it et al.} (D\O\ Collaboration),
  Nucl. Instrum. Methods A {\bf 338}, 185 (1994).

\bibitem{HERWIG}
            G. Marchesini and B.R. Webber, Nucl. Phys. {\bf B310}, 461 (1988).
\bibitem{PYTHIA}  H.-U. Bengtsson and T. Sj\"ostrand, Comp.
 Phys. Comm. {\bf 46}, 43 (1987).

\bibitem{bob} R.S. Fletcher, private communication and verified
with further study by the present authors.

\bibitem{GLASGOW} A. Brandt, {\it Proceedings of the 27th
             International Conference on High
             Energy Physics}, (1994), ed. P.J. Bussey, I.G. Knowles
             (Institute of Physics Publishing, 1995).

\end{references}
\end{document}